\documentclass[aps,pra,superscriptaddress]{revtex4}
\usepackage[T1]{fontenc}
\usepackage[utf8]{inputenc}
\usepackage[english]{babel}
\usepackage{amsmath,amssymb,amsfonts}
\usepackage{graphicx,graphics}
\usepackage{subfig}
 \usepackage{caption}
 \usepackage{dcolumn}
\usepackage{bm}
\usepackage{color}
\usepackage{amsthm}             %para el teorem style
\usepackage{amssymb}
\usepackage{mathrsfs}
\usepackage{dcolumn}% Align table columns on decimal point
\usepackage{bm}% bold math
\usepackage{epsfig}
\usepackage{epstopdf}
\usepackage{float}

\begin{document}

\title{Spontaneous transition rates near the focus of a parabolic mirror with identification of the vectorial modes involved}

\author{R.~Guti\'{e}rrez-J\'{a}uregui}
\affiliation{Department of Physics, Columbia University, New York, NY, USA.}
\email{r.gutierrez.jauregui@gmail.com}
\author{R. J\'auregui}
\affiliation{Instituto de F\'{\i}sica, Universidad Nacional Aut\'onoma de M\'exico, Apdo. Postal 20-364, 01000 Cd. de M\'exico, M\'exico}
\email{rocio@fisica.unam.mx}

\begin{abstract}
{Each natural mode of the electromagnetic field within a parabolic mirror exhibits spatial localization and polarization properties that can be exploited for the quantum control of its interaction with atomic systems. The region of localization is not restricted to the focus of the mirror leading to a selective response of atomic systems trapped on its vicinity. We report calculations of the spontaneous emission rates for an atom trapped inside the mirror accounting for all atomic polarizations and diverse trapping regions. It is shown that electric dipole transitions can be enhanced near the focus of a deep parabolic mirror with a clear identification of the few vectorial modes involved. Out of the focus the enhancement vanishes gradually, but the number of relevant modes remains small. Ultimately this represents a quantum electrodynamic system where internal and external degrees of freedom cooperate to maximize a selective exchange and detection of single excitations.}
\end{abstract}

\maketitle

\section{Introduction}

The theory of spontaneous emission is centered on the idea that an emitter interacts with the surrounding electromagnetic (EM) environment, constantly probing its spatial and spectral structure. Thus, the emission rate can be enhanced~\cite{Purcell_1946} or inhibited~\cite{Kleppner_1981} by properly tailoring the environment, as is widely accepted due to the results of beautiful experiments across different platforms~\cite{Gabrielse_1985,Hulet_1985,Qdtions,Goy_1983,Jhe_1987}. With recent developments in the study of structured light~\cite{structured} and the fabrication of new cavity geometries~\cite{leuchs}, attention now turns from a controlled emission rate to the possibility of selective emission.

The advent of trapped ion physics~\cite{Gabrielse_1985,Hulet_1985,Qdtions}, cavity~\cite{Goy_1983,Jhe_1987} and circuit electrodynamics~\cite{Wallraff_2004} have guided the exploration of these phenomena by imposing different boundary conditions over field and emitter. Naturally, each architecture presents advantages and challenges of its own. Take the case of electrons trapped inside a microwave cavity~\cite{Gabrielse_1985,Hulet_1985}. Here the emission rate can be manipulated by restricting the spectral mode density inside the cavity, but, due to the apertures required to control the electronic motion, the electron still interacts with a large number of spatial modes. This sets a lower limit on the emission rate and introduces challenges on the detection of the emitted photon~\cite{Parkins_1993}. By comparison, superconducting circuits allow for a reduced  number of available spatial modes and offer enviable control over the vacuum fluctuations~\cite{Toyli_2016}. This paves a way to explore the effect of engineered environments over spontaneous emission~\cite{Toyli_2016,Gardiner_1986,Carmichael_1987}, but, given the macroscopic character of the artificial atom, probing regions of large field gradients remains a challenge.

Atomic systems trapped inside parabolic mirrors provide an interesting balance between these two schemes. This half-cavity arrangement allows for detection over nearly the $4\pi$ solid angle surrounding the atom~\cite{leuchs,Maiwald_2009,wang_2020}, and opens the possibility to identify novel effects introduced by the atomic centre-of-mass degrees of freedom. Theoretical studies for quantum electrodynamical processes inside parabolic mirrors, which led to insightful ideas on photon scattering~\cite{Stobinska_2012,Alber_2013}, were, however, deterred due to mathematical difficulties in finding a complete set of modes that fulfill Maxwell equations and the adequate boundary conditions. The problem was surpassed on Ref.~\cite{us} where a detailed description of the electromagnetic field inside parabolic geometries was reported. The natural parabolic EM modes exhibit a non-trivial and rich topology: regions where electric and magnetic field have different magnitude, phase singularities, vectorial vortices, and strong gradients of the field components.

Here, we discuss the spatial properties of parabolic modes with emphasis on their strong localization and describe the profound effect it has over the spontaneous emission of an ion trapped near the focal axis of a deep parabolic mirror.
It is shown that for trapping potentials centered near the focal point an enhanced transition rate can be attributed to only a handful of EM modes. For an atom polarized along the axis, nearly one third of this rate can be attributed to the emission into a single parabolic mode. This result is consistent with recent experimental observations~\cite{Salakhutdinov_2020}. The  behaviour is shown to extend for an atom polarized perpendicular to the axis and for an atom outside the focus. In these cases the number of relevant modes increases but remains small.
The possibility to control the interaction between an atom and the EM field by manipulating few modes in a scenario that allows for a nearly $4\pi$ detection of the field, opens the doors for novel studies of quantum and classical fluctuations in atomic systems. 

\section{Natural modes of the electromagnetic field within a parabolic mirror}

Vectorial parabolic modes~\cite{us} define the natural basis to describe the EM field inside a parabolic mirror. These modes are characterized by a set of parameters denoted by $\gamma$ and comprised of: (i) the frequency $\omega$ linked to the energy of the field in the quantum realm; (ii) an integer number $m$ linked to the total (orbital plus helicity) angular momentum  projected along the symmetry axis of the mirror ($z$-axis); (iii) a parameter $\kappa$ ubiquitous in parabolic geometries that is linked to the cross product of linear and total angular momentum projected along the $z$-axis. In a similar way to the EM field in waveguides, the  polarization behavior of the EM modes with  parabolic symmetry is encoded in the possibility of writing either the electric or magnetic field as the curl of a vectorial Hertz potential. The corresponding modes are referred to as $\mathcal{E}$- and $\mathcal{B}$-modes. That is,
$\gamma=\{\omega,m,\kappa, \mathcal{P}\}$, where $\mathcal{P}$ can be either $\mathcal{E}$ or $\mathcal{B}$.

The electric field  of a parabolic $\gamma$-mode,  ${\boldsymbol{E}}^{(\gamma)}$, can be written in terms of simple trigonometric functions in the restricted Fourier space where the dispersion relation $\vert \mathbf{k}\vert^2 =\omega^2/c^2$ has been taken into account,
\begin{equation}
{\boldsymbol{E}}^{(\gamma)}(\mathbf{x},t)= \int d\Omega_{\mathbf k} e^{i(\omega/c)(\hat{\mathbf{k}}\cdot \mathbf{r}- ct)} \mathbf{f}^{(\gamma)}(\theta_{\mathbf{k}},\varphi_{\mathbf{k}}).
\end{equation}
Since the Fourier representation of the Hertz potential is
\begin{equation}
{\boldsymbol{\pi}}^{(\gamma)}(\mathbf{x},t) =
\sum_{\sigma}\int d\Omega_{\mathbf k} \mathbf{e}_\sigma e^{i(\omega/c)(\hat{\mathbf{k}}\cdot \mathbf{r}- ct)} \tilde{\pi}^{(\gamma)}_\sigma(\theta_{\mathbf{k}},\varphi_{\mathbf{k}}),\label{eq:angular_spectrum}
\end{equation}
\begin{equation}
\tilde{\pi}^{(\gamma)}_\sigma(\theta_{\mathbf{k}},\varphi_{\mathbf{k}})=
\frac{e^{i(m-\sigma)\varphi_{\mathbf{k}}}}{2\pi}\sum_{\kappa_\sigma} \tilde{c}^\sigma_{\kappa_\sigma}  \frac{(\tan\theta_{\mathbf{k}}/2)^{-2\kappa_\sigma i}}{\sin\theta_{\mathbf{k}}},\label{eq:pisigma}
\end{equation}
then $\mathbf{f}^{(\gamma)}=i\omega/c\hat{\mathbf{k}}\times\tilde{\boldsymbol{\pi}}^{(\gamma)}$ for $\mathcal{E}$-modes and $\mathbf{f}^{(\gamma)}=i(\omega/c)\hat{\mathbf{k}}\times(\hat{\mathbf{k}}\times\tilde{\boldsymbol{\pi}}^{(\gamma)})$
for $\mathcal{B}$-modes. We have adopted the circular polarization basis $\{\mathbf{e}_\sigma\}= \{\mathbf{e}_\pm = \mathbf{e}_x \pm i\mathbf{e}_y, \mathbf{e}_0 = \mathbf{e}_z\}$ to highlight the coupling between $\sigma$ and $m$ into an effective winding number $n = m -\sigma$ for each component of the potential ${\boldsymbol{\pi}}^{(\gamma)}$. This structure is inherited to $\mathbf{f}^{(\gamma)}$. The coefficients $\tilde{c}^\sigma_{\kappa_\sigma}$ are constants with values restricted by the symmetries, boundary conditions and, for quantum fields, normalization conditions \cite{us}.  

\begin{figure}
\includegraphics[width=.65\linewidth]{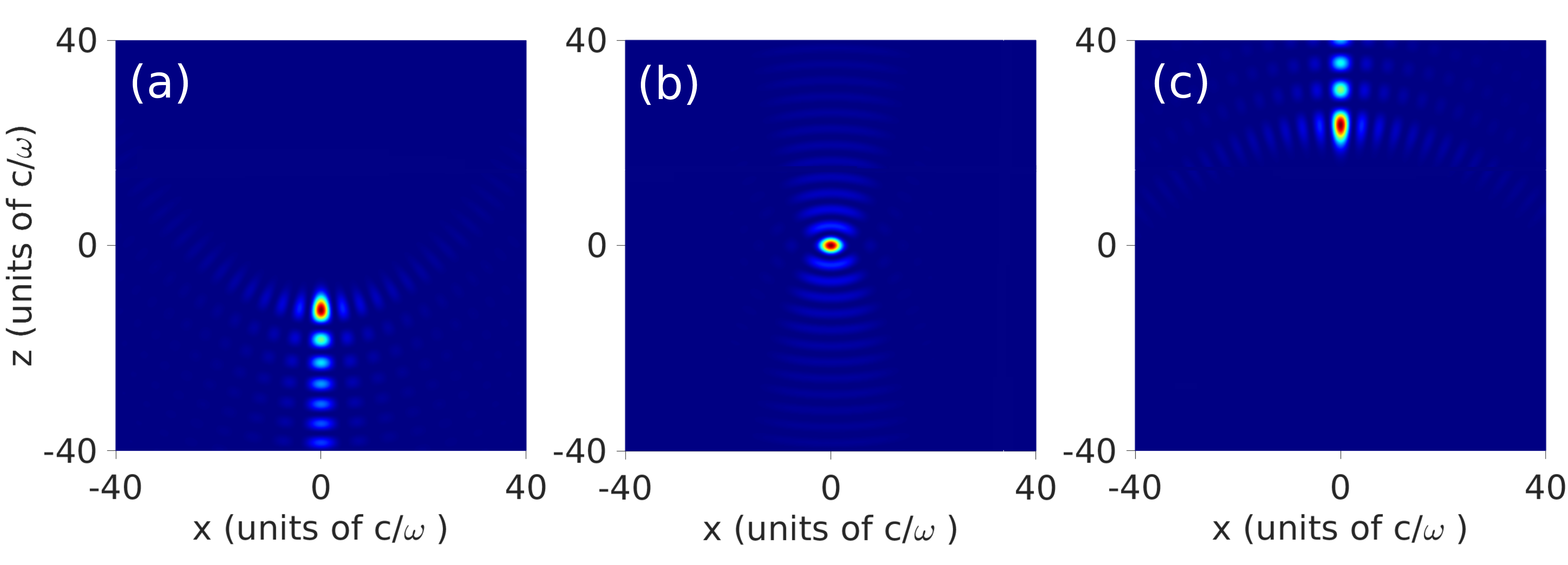}\\
\includegraphics[width=.65\linewidth]{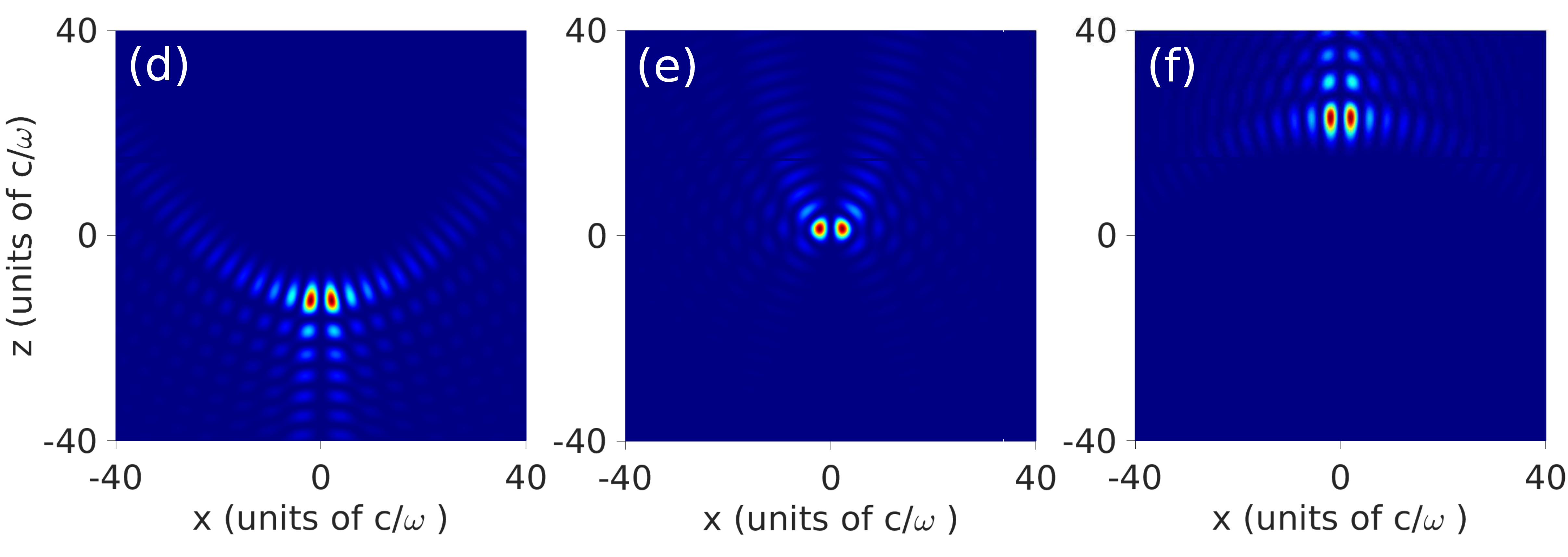}\\
\includegraphics[width=.65\linewidth]{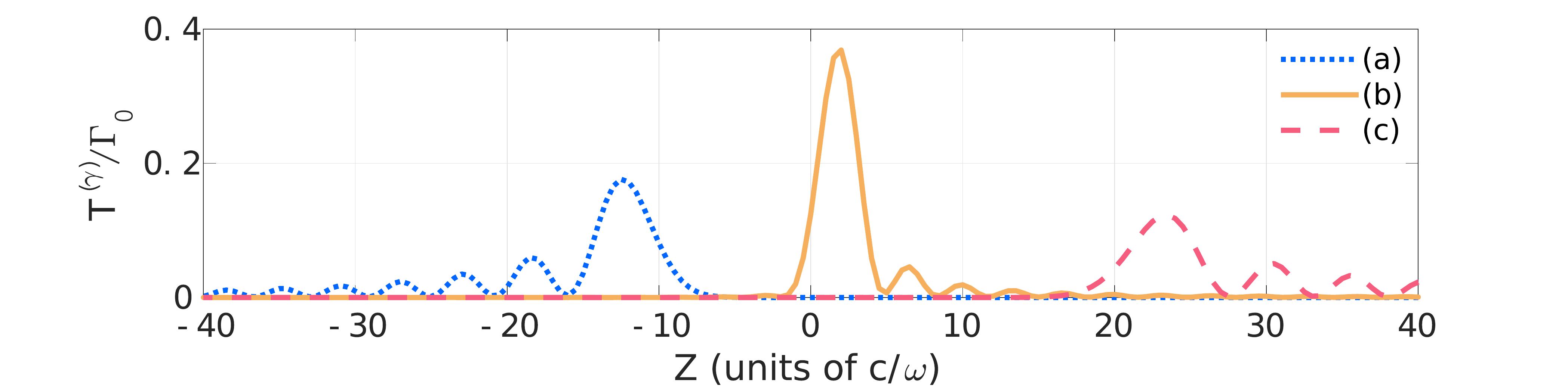}\\
\caption{Relative intensity of  $\boldsymbol{E}_z$ for $\mathcal{E}$-modes with $m= 0$ (a-c) and $\vert m\vert =1$ (d-f). In (a) and (d) $\kappa = 5.6$, (b) and (e) $\kappa =-0.64$, (c) and (f)
$\kappa = -10.4$. 
The third row illustrates the contribution of individual  $\mathcal{E}$-modes (a-c) to the spontaneous emission rate of a dipole parallel to the $z$-axis in a mirror with a focal distance about $71500 c/\omega$. $\Gamma_0$ refers to the transition rate in free space.}\label{fig:localization}
\end{figure}

All parabolic vectorial $\gamma$-modes exhibit a high degree of spatial localization, displaying narrow regions of maximum intensity near a plane defined by a particular value $Z$ of the $z$-coordinate. The plane is determined by the parameter $\kappa$, as readily inferred from the angular spectrum $\mathbf{f}^{(\gamma)}$. 
From Eq.~(\ref{eq:pisigma}), all $\sigma$-components of  the electric field involve integrals
\begin{equation}
R^{(\gamma)}_\sigma(\rho,Z)=\int_0^{2\pi} d\theta_{\mathbf{k}} e^{i(Z\omega/c)\cos\theta_{\mathbf{k}}-i2\kappa\log(\tan\theta_{\mathbf{k}}/2)}
h^{(\gamma)}_\sigma(\theta_{\mathbf{k}};\rho)
\end{equation}
with a rapidly oscillating phase term (for $\vert Z\omega/c\vert\gg 1$) and  a function $h^{(\gamma)}_\sigma$ that depends on the distance $\rho$ to the $z$-axis, but not  on $Z$.
Using the stationary phase method
\begin{equation}
R^{(\gamma)}_\sigma(\rho,Z)\simeq \sqrt{\frac{\pi}{\vert Z\omega/c\vert \sqrt{1 + 2\kappa c/Z\omega}}}h_\sigma^{(\gamma)}(\tilde\theta_{sp};\rho),\label{eq:rstph}\end{equation}
such that the maximum amplitude is achieved near the plane  $Z\sim-2\kappa c/\omega$. The stationary phase condition $\sin^2\tilde\theta_{sp} =-2c\kappa/Z\omega$ is feasible only if $0\le -2c\kappa/Z\omega\le 1$. Figure~\ref{fig:localization} illustrates the localization of several parabolic modes. The electric field components are evaluated numerically  without using the stationary phase approximation. 
For a component of the field with $n=0$ the maxima is on the $z$-axis, while for a mode exhibiting a vortex on axis, the maximum  amplitude is expeled to a ring out of the $z$-axis but remains nearby the plane.

\section{Spontaneous decay of an atom trapped inside a parabolic mirror}

The localization of the modes directly influences the response of an atom to the EM field.
An atom trapped inside the mirror couples to the parabolic modes via the interaction Hamiltonian which, in the dipole approximation, is given by 
\begin{equation}
\widehat{H}_{int} =- \widehat{\mathbf d}\cdot \widehat{\mathbf E}({\mathbf X},t),
\end{equation}
with $\widehat{\mathbf{d}} = e\widehat{\mathbf{r}}$ the electric dipole moment operator of the electron of charge $e$ which changes its  orbital  in the transition, and  $\widehat{\mathbf E}$  the electric field  operator evaluated at  the atomic center-of-mass (CM) coordinate ${\mathbf{X}}$ at time $t$.  $\widehat{\mathbf E}$
can be written as superposition of the creation and annihilation operators of each mode $\gamma$  with c-number coefficients given by the corresponding electric field amplitude  ${\boldsymbol{E}}^{(\gamma)}$.

We focus on the spontaneous decay process. The transition matrix connecting an initial electronic state $\vert a\rangle$ with bounded CM state $\vert \Phi_{A}\rangle$ to a final state $\vert b\rangle$ with CM state $\vert \Phi_{B}\rangle$ through the emission of a photon $\gamma$ is then
\begin{equation}
 M_{ab;AB}^{(\gamma)} = \frac{\omega_{ab}}{\omega_{\gamma}} \mathbf{d}_{ab} \cdot \left[\int \text{d}^{3}\mathbf{X}  \Phi_{A}^{*}(\mathbf{X}) {\boldsymbol{E}}^{(\gamma)*}(\mathbf{X}) \Phi_{B}(\mathbf{X})  \right] e^{i (\omega_{ab}+\omega_{AB}-\omega_{\gamma})t}
\end{equation}
where $\omega_{ab}$ and $\omega_{AB}$ are the electronic and vibrational transition frequencies, and $\omega_{\gamma}$ the mode frequency. This transition matrix  determines one of the many paths the atom can follow to reach the lower state. The probability to follow each path depends on the cavity mode structure and available vibrational states and is weighted by the coupling strength. We remain in the perturbative limit of CQED, and consider the electronic transition frequency to be much larger than the relevant CM transition frequencies $(\omega_{ab} \gg \omega_{AB})$. Under these conditions the spontaneous transition rate between the internal atomic states, irrespective to the final CM motion and emitted photon mode $\gamma$ is
\begin{eqnarray}
\Gamma_{ab}^A &=&\frac{2\pi}{\hbar^2}\sum_{B,\gamma}{\mathcal{S}}(\omega_{ab}+\omega_{AB}-\omega_\gamma)
\bigg|{\mathbf d}_{ab}\cdot\int d^3\mathbf{X}\boldsymbol{E}^{(\gamma)}(\mathbf{X})
\Phi^*_A(\mathbf{X})\Phi_B(\mathbf{X})\bigg| ^2 \nonumber\\
&\approx&\frac{2\pi}{\hbar^2}\sum_{\gamma,j,k} {\mathcal{S}}(\omega_{ab}-\omega_\gamma)({\mathbf{d}}_{ab})^*_k
 ({\mathbf d}_{ab})_j\int d^3\mathbf{X}\Phi^*_{A}(\mathbf{X}){\boldsymbol{E}}^{(\gamma)*}_k(\mathbf{X})
 {\boldsymbol{E}}^{(\gamma)}_j(\mathbf{X})\Phi_{A}(\mathbf{X}).\label{eq:aprox}
\end{eqnarray}
with $\mathcal{S}$ a sharp spectral function centered at the internal transition frequency resulting from the temporal integration of the interaction. Written in this form, the spontaneous emission rate becomes proportional to the autocorrelation function of the electric field averaged over the initial CM position~\cite{Jauregui_2015,Trautmann_2016}. This opens the possibility of incorporating information about the spatial region explored by the atom in the calculation of $\Gamma_{ab}^A$ in a modular way.  Using the angular spectrum representation of the electric field, $\mathbf{f}^{(\gamma)}$,
\begin{equation}
\Gamma_{ab}^A =\sum_{\sigma=,\pm,0}\sum_{\gamma}{\mathcal{S}}(\omega_{ab}-\omega_\gamma)\vert({\mathbf{d}}_{ab})_{-\sigma}\vert^2\mathcal{T}_{{A};\sigma}^{(\gamma)},\label{eq:TIF:IIF}
\end{equation}
where the contribution of each mode is given by,
\begin{eqnarray}
\mathcal{T}_{{A};\sigma}^{(\gamma)}&=&\frac{2\pi}{\hbar^2} \int d\Omega_{\mathbf k} \int d\Omega_{\mathbf{k}^\prime} 
\mathbf{f}^{(\gamma)*}_\sigma(\theta_{\mathbf{k}},\varphi_{\mathbf{k}})g(\mathbf{k},\mathbf{k}^\prime;\omega_\gamma;A)
\mathbf{f}^{(\gamma)}_\sigma(\theta_{\mathbf{k}^\prime},\varphi_{\mathbf{k}^\prime}),\label{eq:TIF}\\
g(\mathbf{k},\mathbf{k}^\prime;\omega_\gamma;{A})&=& \int d^3\mathbf{X}\Phi^*_{{A}}(\mathbf{X})e^{i(\hat{\mathbf{k}}-\hat{\mathbf{k}}^\prime)\cdot(\omega_\gamma \mathbf{X}/c)}
 \Phi_{{A}}(\mathbf{X}).\label{eq:ff}
\end{eqnarray}
The scalar form factor $g(\mathbf{k},\mathbf{k}^\prime;\omega_\gamma;A)$ incorporates all the information about the initial CM state $\vert A\rangle$.  Since parabolic modes are highly localized, trapping the atom nearby a given spatial region can guarantee that the atom interacts only with a finite number of modes. In Fig.\ref{fig:localization}, the contribution to the transition rate of a single mode, $\mathcal{T}_{A;\sigma}^{(\gamma)}$ 
is illustrated for different $\mathcal{E}$-modes for an atom tightly trapped at the position $\mathbf{X}$ centered at the $z$-axis. The modes are normalized so that their electric field corresponds to that of a single photon \cite{us}. It is important to remark that their contribution reaches its maximum value at $Z \sim -2\kappa c/\omega$ with little overlap between different modes. 

The spatial distribution of a given mode competes with the symmetry properties introduced by the trapping potential. Consider an atom trapped in a harmonic potential centered at the position $\mathbf{X}_0$ which does not necessarily coincide with the origin used to describe the EM field. Then, the form factor can be written as
\begin{equation}
g(\mathbf{k},\mathbf{k}^\prime;\omega_\gamma;0) = e^{i(\hat{\mathbf{k}}-\hat{\mathbf{k}}^\prime)\cdot(\omega_\gamma \mathbf{X}_0/c)}
\prod_{i=x,y,z} e^{-\frac{{\boldsymbol{\eta}}^{2}_{i}(\hat{k}_i -\hat{k}_i^\prime)^2}{2}}\label{eq:ffho0}
\end{equation}
for an atom in the ground state of the trap; here the trap frequencies $\{\boldsymbol{\Lambda}_x,\boldsymbol{\Lambda}_y, \boldsymbol{\Lambda}_z\}$ are incorporated via the Lamb-Dicke parameters ${\boldsymbol{\eta}}_{i} =\sqrt{\hbar\omega^2_\gamma/2M\boldsymbol{\Lambda}_i c^2}$, with $M$ the atomic mass. If the trap is symmetric under rotations around the $z$-axis, that is ${\boldsymbol{\eta}}_{x} = {\boldsymbol{\eta}}_{y}$, then the form factor depends on $\varphi_{\mathbf{k}}$ and $\varphi_{\mathbf{k}^\prime}$ only through their difference. 
Introducing $g(\mathbf{k},\mathbf{k}^\prime;\omega_\gamma;0)$ in Eq.~(\ref{eq:TIF}), and taking $\mathbf{X}_0 = (0,0,Z)$, allows for the azimuthal integral to be solved  
\begin{equation}
\int_{-\pi}^\pi d\varphi_{\mathbf{k}^\prime}\int_{-\pi}^\pi d\varphi_{\mathbf{k}} 
e^{-i(n^\prime\varphi_{\mathbf{k}^\prime} -n\varphi_{\mathbf{k}})} 
e^{(\boldsymbol{\eta}_x)^2(\sin\theta_{\mathbf{k}}\sin\theta_{\mathbf{k}^\prime} 
\cos(\varphi_{\mathbf{k}}-\varphi_{\mathbf{k}^\prime}))}=
(2\pi)^2\delta_{n^\prime,n}I_{\vert n\vert}\Big(\vert(\boldsymbol{\eta}_{x})^2\sin\theta_{\mathbf{k}}\sin\theta_{\mathbf{k}^\prime}\vert\Big)\label{Eq;In}
\end{equation}
with $I_{\vert n\vert}$ a modified Bessel function of index $n=m-\sigma$ [see Eq.~(\ref{eq:pisigma}) above]. Since  $I_{\vert n+1\vert}(x) <I_{\vert n\vert}(x)$ for $x>0$ with a wider separation as its argument increases, the leading contribution to the spontaneous decay is given by $n=0$ terms and becomes more dominant as the axial confinement rises. In the limit of infinite trap frequency $I_{\vert n\vert} \rightarrow \delta_{n0}$.

The presence of an ideal parabolic mirror constraints the available parabolic modes depending on the transition frequency \cite{us}. To study this effect, we consider the $^2S_{1/2}\rightarrow$ $^2P_{1/2}$ transition for YbII  and YbIII  ions following the experimental parameters reported in Reference~\cite{trap}. The ion is assumed to be in the motional ground state of a radiofrequency Paul trap with radial secular frequencies of 230 KHz and 460 KHz, and axial secular frequencies of 480KHz and 960 KHz for YbII and YbIII, respectively. 

\begin{figure}[hb]
\centering
\includegraphics[width=.5\columnwidth]{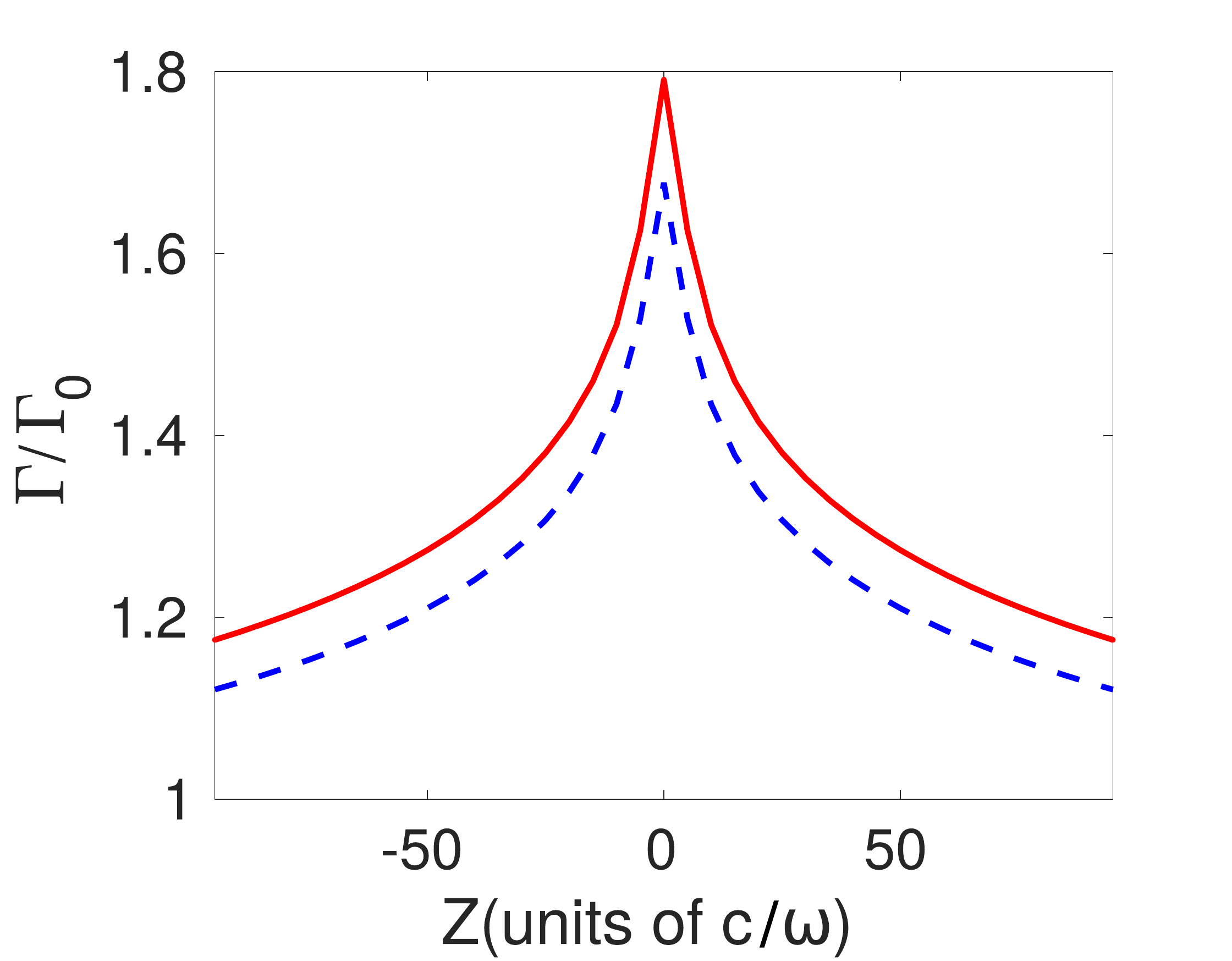}
\caption{Dependence of the spontaneous transition rate $\Gamma$ on the distance between trap center and focal point for YbII ion (red solid) and  YbIII ion (blue dashed). The ion is polarized along the $z$-axis with parameters specified in the main text. }\label{fig:TIF:Neu}
\end{figure}

\begin{figure}[ht]
 \includegraphics[width=1.\columnwidth]{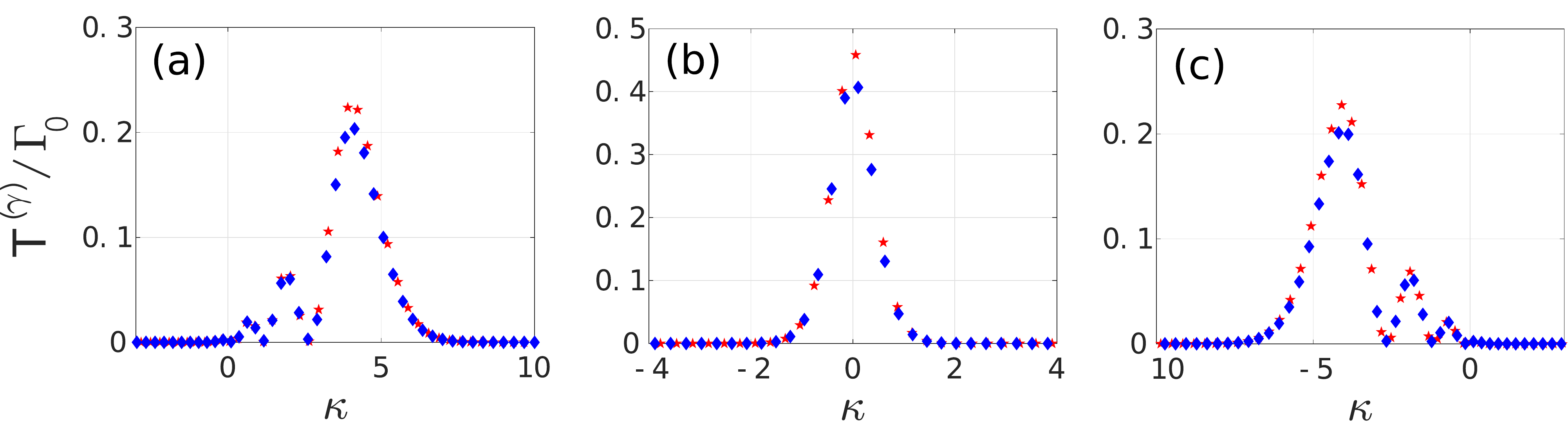}
\caption{Individual contribution of ${\mathcal E}$-modes with $m=0$ to $\Gamma$ for an  YbII ion (red stars) and  YbIII ion (blue diamonds) polarized along the $z-$axis and located at: (a)  $Z =-10 c/\omega$; (b) $Z = 0$; (c)  $Z =10 c/\omega$. The parameters are the same of Fig.~\ref{fig:TIF:Neu}.}\label{fig:neu:modes}
\end{figure}

\begin{figure}[ht]
\centering
 \includegraphics[width=.6\columnwidth]{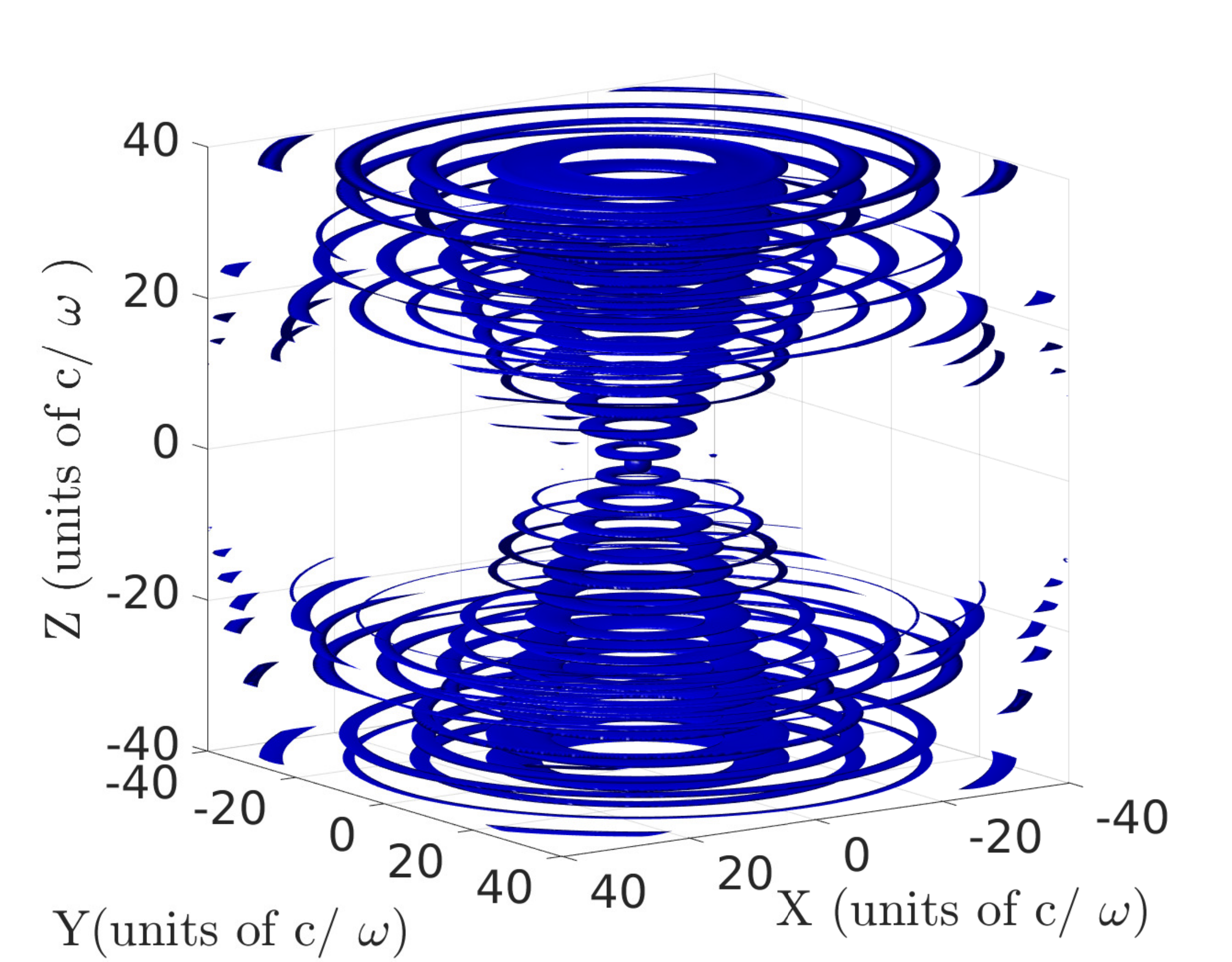}
\caption{Isointensity suface of a ${\mathcal E}$-mode with $m=0$ and $\kappa= 0.02$.  On this surface the electric field modulus equals $\sim$0.42 its maximum value.}\label{iso:neu:kappazero}
\end{figure}

For an atomic dipole moment oriented along the $z$-axis coinciding with the mirror, quantization, and trap axes, the dominant modes are expected to be the $\mathcal{E}$-modes with  $m=0$.  In Fig.~\ref{fig:TIF:Neu} the resulting spontaneous decay rate is illustrated for both YbII  and YbIII. This rate is larger than the free space value $\Gamma_0$ for the atom located in a region  within a $\sim 25 c/\omega$ range around the focus of the mirror (focus distance 2.1mm). This interesting result is supplemented by knowledge of the contribution of each parabolic mode. In Fig.~\ref{fig:neu:modes} the $\mathcal{T}_{A;\sigma}^{(\gamma)}$ values for the individual significant modes are shown. At the focus of the mirror a single mode with $\kappa\sim 0.02$ contributes with $\mathcal{T}_{A;\sigma}^{(\gamma)}\sim 0.47\Gamma_0$ for YbII, meaning that this mode has a probability of nearly one third to be populated each time the atom decays. The maximum intensity of that mode is located almost exactly at the focus of the mirror. The structure of the complete local density of energy of the mode with $\kappa= 0.02$ as described by isosurfaces is illustrated in Fig.~\ref{iso:neu:kappazero}. Its similarity with the expected radiation pattern of a classical oscillating dipole can be observed. 
The other dominant contributions correspond to modes with $\vert \kappa\vert < 2$. Due to the small differences between spatial profiles of these dominant modes, a standard detector might not be able to distinguish between different but nearby values of $\kappa$. By comparison, the minimum value of $\vert \kappa\vert$ allowed by the boundary condition for YbIII corresponds to $\kappa\sim 0.64$ and yields  $\mathcal{T}_{A;\sigma}^{(\gamma)}\sim 0.4\Gamma_0$ for $Z=0$. This exemplifies the dependence on the light frequency of the allowed modes within an ideal parabolidal mirror. It also reiterates that modes whose adequate field component has a maximum at the position of highest probability for the atom are more likely to be emitted [see Fig.~\ref{fig:localization}].  
Outside the focus the number of significant modes increases but it is always small, and in this case is $\sim 10$. Eventhough the Paul trap yields higher trapping frequencies for YbIII than for YbII, the number of significant modes is similar for both atoms  as expected from the similarity of their Lamb-Dicke parameters.

\begin{figure}[hb]
\centering
 \includegraphics[width=.5\columnwidth]{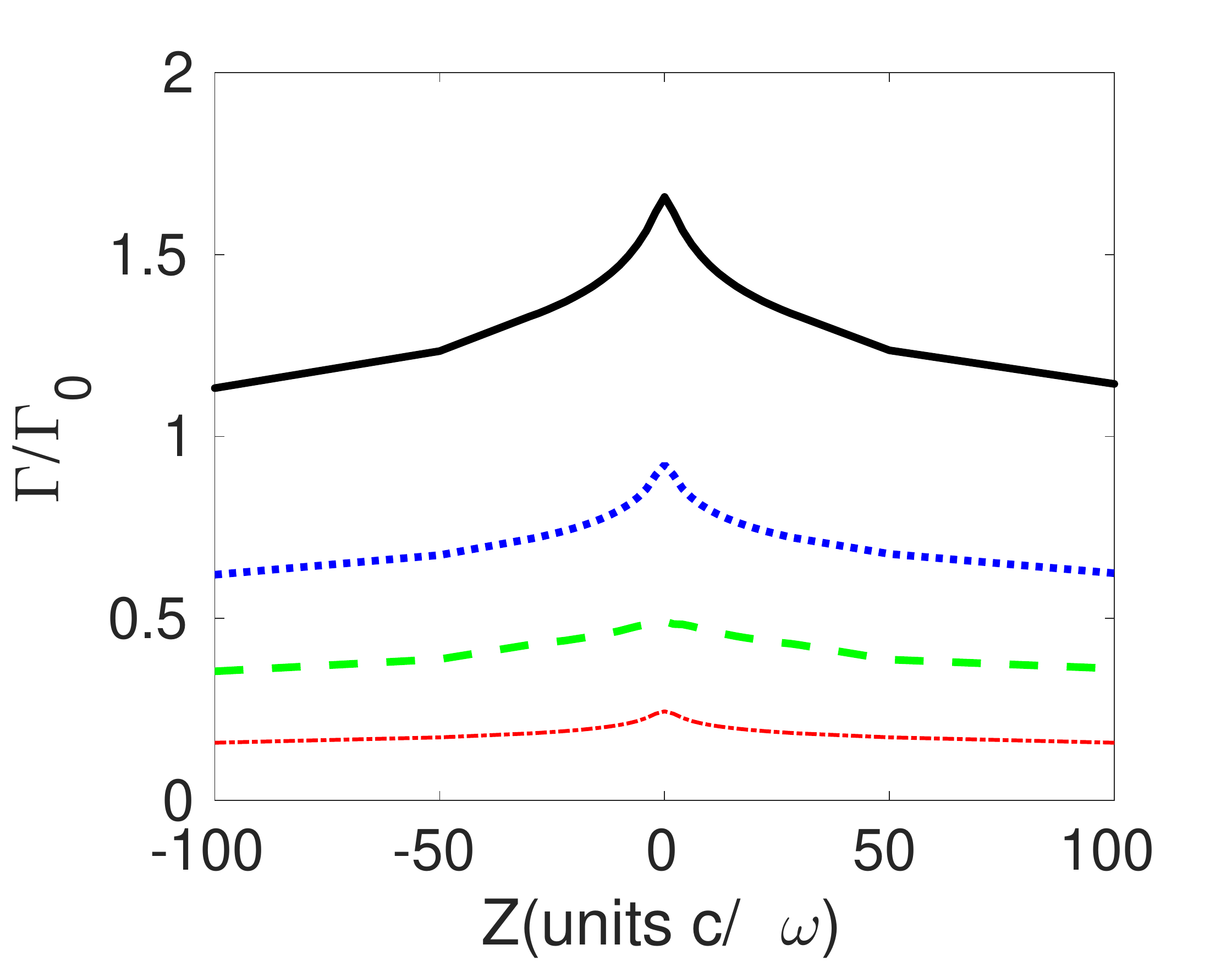}
\caption{Spontaneous transition rate (black solid) for an YbII ion polarized perpendicular to the $z-$axis. It results from adding the individual contribution of the $\mathcal{E}$-modes with $\vert m\vert =1$ (blue dashed), $\mathcal{B}$-modes with $\vert m\vert =1$ (green long-dashed) and  $\mathcal{B}$-modes with $m=0$ (red dot-dashed). The general parameters are the same of Fig.~\ref{fig:TIF:Neu}.}\label{trans-perp}
\end{figure}

For dipole moments perpendicular to the mirror axis the atom probes different field components. As already mentioned, just the ${\mathcal E}$ and ${\mathcal B}$ modes derived from Hertz potentials with $\vert m\vert =1$ can yield $n=0$ , and thus satisfy the relevance condition of Eq.~(\ref{Eq;In}).  The spatial distribution of the electric field for these modes has been illustrated in Ref.~\cite{us}. A direct calculation using individual modes with the parameters previously described is consistent with these expectations as illustrated in Fig.~\ref{trans-perp} where the total spontaneous decay (black) and the main contributions are plotted. The ${\mathcal B}$-modes with $\vert m\vert=1$ are emitted with a higher probability than those with $m=0$. Similarly to the case of a dipole parallel to the $z$-axis, the total spontaneous decay rate $\Gamma$ at the focus of the parabola is enhanced, 
and $\Gamma \rightarrow \Gamma_0$ as the atom leaves the mirror focus. For each kind of modes and atom position, a close range of $\kappa$ values yield significant contributions to $\Gamma$.

The panorama changes if the atom is located outside the $z$-axis. In this case the probabilty of emission of modes with a given $m$ will depend on the atomic distance to the mirror axis.  Each component of ${\boldsymbol{E}}^{(\gamma)}$ remains strongly localized (at a $z$-position determined by $\kappa$), but the distance between this maximum and the symmetry axis depends on the effective winding number $n$ as illustrated in Fig.~\ref{fig:localization}.  This feature can prove advantageous to control the vorticity of the  emitted light  by locating the atom at the adequate distance to the optical axis.

\section{Discussion}

It has been shown that the spontaneous decay of atomic systems located on the axis of ideal parabolic mirrors involve a compact set of electromagnetic modes that can be identified by their space localization and local polarization properties. For a detector unable to distinguish between modes with different, but nearby values of $\kappa$, the quantum correlation functions of the emitted light by an atom situated on the axis of the mirror are almost identical to those expected from a system that just interacts with a single mode. 

The formalism described here can be generalized to a driven scenario where the modes are populated by an external source using, for instance, a space light-modulator. When those modes impinge on a localized atom, the scattering pattern could inherit the peculiar spatial and polarization structure of the incident light.
If a parabolic surface is properly included in an experimental set up the symmetry effects are enhanced. In this scheme, the atom would scatter the modes with a given $\gamma$ parameters with a strong dependence on the atom spatial localization about the mirror focal point.

Our calculations predict that atomic spontaneous decay can be enhanced about a 75$\%$  for Yb ions close to the focus of an ideal macroscopic parabolic mirror. Smaller mirrors, or atoms located close to the surface of a mirror with high curvature could yield new interesting phenomena. The spatial distribution of energy, orbital angular momentum and polarization of EM parabolic modes, as well as the open character of the parabolic boundary, are ideal for the performance of novel quantum optics experiments where many degrees of freedom can be accessed and manipulated to control the atomic state.

\section*{Acknowledgements}
This work was partially supported by PAPIIT-UNAM 103020 and PIIF-UNAM-08-2019.

%\section*{Author contributions}
%RGJ and RJ contributed with conceptual ideas, methodology strategies and writing of the manuscript. Numerical calculations were performed by RJ.

%\section*{Competing interests}
%The authors declare no competing interests.

\end{document}